\documentclass
[10pt,letterpaper,bibnotes,notitlepage,final,balancelastpage,superscriptaddress,twocolumn,showpacs,pra]{revtex4}%
\usepackage{amssymb}
\usepackage{amsmath}
\usepackage{amsfonts}
\usepackage{graphicx}%
\begin{document}
\title{Fault-tolerant quantum repeater with atomic ensembles and linear optics}
\author{Zeng-Bing Chen}
\affiliation{Hefei National Laboratory for Physical Sciences at Microscale and Department
of Modern Physics, University of Science and Technology of China, Hefei, Anhui
230026, China}
\author{Bo Zhao}
\affiliation{Physikalisches Institut, Universit\"{a}t Heidelberg, Philosophenweg 12,
D-69120 Heidelberg, Germany}
\author{Yu-Ao Chen}
\affiliation{Physikalisches Institut, Universit\"{a}t Heidelberg, Philosophenweg 12,
D-69120 Heidelberg, Germany}
\author{{J\"{o}rg Schmiedmayer}}
\affiliation{Atominstitut der \"{O}sterreichischen
Universit\"{a}ten, TU-Wien, A-1020 Vienna, Austria}
\author{Jian-Wei Pan}
\affiliation{Hefei National Laboratory for Physical Sciences at
Microscale and Department of Modern Physics, University of Science
and Technology of China, Hefei, Anhui 230026, China}
\affiliation{Physikalisches Institut, Universit\"{a}t Heidelberg,
Philosophenweg 12, D-69120 Heidelberg, Germany}


\begin{abstract}
We present a detailed analysis of a new robust quantum repeater
architecture building on the original DLCZ protocol [L.M. Duan
\textit{et al.}, Nature (London) \textbf{414}, 413 (2001)]. The
new architecture is based on two-photon Hong-Ou-Mandel-type
interference which relaxes the long-distance interferometric
stability requirements by about 7 orders of magnitude, from
sub-wavelength for the single photon interference required by DLCZ
to the coherence length of the photons, thereby removing the
weakest point in the DLCZ schema. Our proposal provides an
exciting possibility for robust and realistic long-distance
quantum communication.

\end{abstract}
\maketitle

\section{introduction}

Quantum communication ultimately aims at absolutely secure
transfer of classical messages by means of quantum cryptography or
faithful teleportation of unknown quantum states
\cite{Gisin,Ekert}. Photons are ideal quantum information carriers
for quantum communication. Unfortunately, photon losses and the
decrease in the quality of entanglement scale exponentially with
the length of the communication channel. The quantum repeater
protocol combining entanglement swapping \cite{marek,swapping} and
purification \cite{bennett} enables to establish high-quality
long-distance entanglement with resources increasing only
polynomially with transmission distance
\cite{repeater1,repeater2}.

Early physical implementations of a quantum repeater were based on
atoms trapped in high-finesse cavities \cite{van}, where strong
coupling between atoms and photons is required. In a seminal paper
\cite{DLCZ}, Duan \textit{et al}. (DLCZ) proposed an
implementation of the quantum repeater by using atomic ensembles
and linear optics. In this protocol atomic ensembles are used as
memory qubits to avoid the challenging request for strong coupling
between atoms and photons. Besides, the DLCZ protocol has built-in
entanglement purification and thus is photon-loss tolerant. In the
efforts of realizing the atomic ensemble based quantum repeater
protocol, significant experimental advances have been achieved
recently. Non-classical correlated photon pairs were generated
from a MOT and a hot vapor \cite{nonclassical1,nonclassical2}.
Controllable single photons were generated from atomic ensembles
with the help of event-ready detection and feedforward circuit
\cite{Eisman,kuzmich,shuai}. Entanglement between two atomic
ensembles either in the same MOT or in two MOTs at a distance of 3
m were generated by detecting single photons
\cite{entanglement,kuzsci}. Recently, segment of the DLCZ protocol
was demonstrated \cite{segment}.

However, the DLCZ protocol has an inherent drawback which is
severe enough to make long-distance quantum communication
extremely difficult. Entanglement generation and entanglement
swapping in the DLCZ protocol depend on a single-photon
Mach-Zehnder-type interference. The relative phase between two
remote entangled pairs is sensitive to path length instabilities,
which has to be kept constant within a fraction of photon's
wavelength. Moreover, entanglement generation and entanglement
swapping are probabilistic. If connecting neighboring entangled
pairs does not succeed after performing entanglement swapping, one
has to repeat all previous procedures to reconstruct the entangled
pairs. This means that the phase fluctuations must be stabilized
until the desired remote entangled pairs are successfully
generated. A particular analysis shows that (see details below),
to maintain path length phase instabilities at the level of
$\lambda/10$ ($\lambda$: wavelength; typically $\lambda\sim1$ $\mu
m$ for photons generated from atomic ensembles) requires the fine
control of timing jitter at a sub-femto second level over a
timescale of a few tens of seconds. This is almost an
experimentally forbidden technique as compared with the lowest
reported jitter in fiber even for kilometer-scale distances
\cite{ye}.

In a recent Letter \cite{zhao}, we proposed a robust quantum repeater
architecture building on the DLCZ protocol . The architecture is based on the
two-photon Hong-Ou-Mandel-type interference \cite{HOM,kimho,kuzmho,yuan},
which is insensitive to phase instability. The path length fluctuations should
be kept on the length scale within a fraction of photon's coherence length
(say, 1/10 of the coherence length, which is about 3 m for photons generated
from atomic ensembles \cite{Eisman}). Therefore the robustness is improved
about 7 orders of magnitude higher in comparison with the single-photon
Mach-Zehnder-type interference in the DLCZ protocol.

In this article, we give a particular analysis on the phase
stability problem in the DLCZ protocol and discuss the robust
quantum repeater architecture in detail. The paper is organized as
follows. In Sec. \uppercase\expandafter{\romannumeral2} we  review
the original DLCZ protocol, and show why the phase stability
problem in the DLCZ protocol is so severe that it makes a
long-distance quantum communication impossible. Sec.
\uppercase\expandafter{\romannumeral3} presents a detailed
analysis of the robust quantum repeater based on the two-photon
Hong-Ou-Mandel-type interference. A comparison with other atomic
ensemble based quantum repeater protocols is also discussed.
Finally, we shall summarize and draw some conclusions.

\section{The DLCZ protocol}

\subsection{Review of the DLCZ protocol}

\begin{figure}
\begin{center}
\includegraphics[height=7cm]{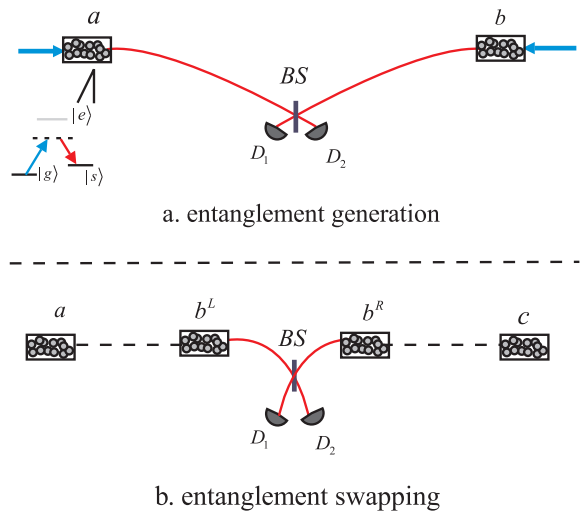}
\end{center}
\caption{Setups for entanglement generation and Entanglement
swapping in the DLCZ protocol. a). Forward-scattered Stokes
photons, generated by an off resonant write laser pulse via
spontaneous Raman transition, are directed to beam splitter (BS)
at the middle point. Entanglement is generated between atomic
ensembles at sites $a$ and $b$, once there is a click on either of
the detectors. The inset shows the atomic level structure, with
the ground state $|g\rangle$, metastable state $|s\rangle$, and
excited state $|e\rangle$. b). Entanglement has been generated
between atomic ensembles $(a,b^{L})$ and $(b^{R},c)$. The atomic
ensembles at site $b$ are illuminated by near resonant read laser
pulses, and the retrieved anti- Stokes photons are subject to BS
at the middle point. A click on either of the detectors will
prepare the atomic ensembles at $a$ and $c$ into an entangled
state.}
\end{figure}

Let us first consider a pencil shaped atomic sample of $N$ atoms with
$\Lambda$ level structure (see inset in Fig. 1). The write laser pulse induces
a spontaneous Raman process, which prepares the forward-scattered Stokes mode
and collective atomic state into a two mode squeezed state \cite{DLCZ}. The
light-atom system can be described as
\begin{equation}
|\psi\rangle=|0_{a}0_{s}\rangle+\sqrt{\chi}S^{\dagger}a^{\dagger}|0_{a}%
0_{s}\rangle
\end{equation}
by neglecting higher-order terms, where $|0_{a}\rangle=\bigotimes_{i}%
|g\rangle_{i}$ is the ground state of the atomic ensemble and $|0_{s}\rangle$
denotes the vacuum state of the Stokes photons. Here, $a^{\dagger}$ is the
creation operator of the Stokes mode, and the collective atomic excitation
operator is defined by $S^{\dagger}=\frac{1}{\sqrt{N}}\sum_{i}|s\rangle
_{i}\langle g|$, where $|s\rangle$ is the metastable atomic state. The small
excitation probability $\chi\ll1$ can be achieved by manipulating the write
laser pulse \cite{thesis}.

The entanglement generation setup is shown in Fig. 1a. Let us consider two
atomic ensembles at site $a$ and $b$ at a distance of $L_{0}\leq L_{att}$,
with $L_{att}$ the channel attenuation length. The Stokes photons generated
from both sites are directed to the beam splitter (BS) at the middle point.
Once there is a click on the detectors, entanglement between communication
sites $a$ and $b$ is established.

The entanglement swapping setup is depicted in Fig. 1b. Assume we
have created entangled states between atomic ensembles ($a,b^{L}$)
and ($b^{R},c$), where $b^{L}$ and $b^{R}$ are at the same site.
The two atomic ensembles at site $b$ are illuminated
simultaneously by read laser pulses. The retrieved anti-Stokes
photons are subject to the BS, and a click on either of the single
photon detectors will prepare the atomic ensembles at sites $a$
and $c$ into an entangled state. The entangled pair can be
connected to arbitrary distance via entanglement swapping.

\subsection{Phase instability analysis I}

In the DLCZ protocol, the single-photon Mach-Zehnder interference
is used in both entanglement generation and entanglement swapping
process. Thus the phase is sensitive to path length fluctuations
on the order of photons' sub-wavelength. Note that to implement
quantum cryptography or Bell inequality detection, one has to
create two pairs of entangled atomic ensembles in parallel
\cite{DLCZ}. The entanglement generated between the two pairs of
atomic ensembles is equivalent to a polarization maximally
entangled state. In this case, the relative phase between the two
entangled pairs needs to be stabilized, which is helpful to
improve the phase instability \cite{segment}. However, the
requirement to stabilize the relative phase in the DLCZ scheme is
still extremely demanding for current techniques.

As shown in Fig. 2, in entanglement generation process the entanglement is
established between the atomic ensembles ($a_{u},b_{u}$) and ($a_{d},b_{d}$)
in parallel during a time interval $t_{0}=\frac{T_{cc}}{\chi e^{-L_{0}%
/L_{att}}}$, where $T_{cc}=L_{0}/c$ is the classical communication time. Note
that one requests $2^{n}\chi\ll1$ to make the overall fidelity imperfection
small, where $n$ is the connection level \cite{DLCZ}. The entanglement
generated between the two pairs of atomic ensembles can be described by
\begin{align}
|\psi_{\phi_{u}}\rangle_{a_{u},b_{u}}  & =(S_{a_{u}}^{\dagger}+e^{i\phi_{u}%
}S_{b_{u}}^{\dagger})/\sqrt{2}|vac\rangle,\\
|\psi_{\phi_{d}}\rangle_{a_{d},b_{d}}  & =(S_{a_{d}}^{\dagger}+e^{i\phi_{d}%
}S_{b_{d}}^{\dagger})/\sqrt{2}|vac\rangle,
\end{align}
where $\phi_{u}=kx_{u}$ ($\phi_{d}=kx_{d}$) denotes the difference of the
phase shifts in the left and the right side of channel $u$ $(d)$, with $x_{u}$
($x_{d}$) the length difference between the left and the right side channel
$u$ $(d)$. Here $k$\ is the wave vector of the photons. For simplicity we have
assumed the lasers on the two communication nodes have been synchronized, and
the phase instability is caused by the path length fluctuations. The
entanglement generated in this process is equivalent to a maximally entangled
polarization state between the four atomic ensembles,
\begin{equation}
|\psi_{\delta\phi}\rangle_{PME}=(S_{a_{u}}^{\dagger}S_{b_{u}}^{\dagger
}+e^{i\delta\phi}S_{a_{d}}^{\dagger}S_{b_{d}}^{\dagger})/\sqrt{2}|vac\rangle,
\end{equation}
where the relative phase between the entangled states of the two
pairs of the remote ensembles is denoted by $\delta\phi=k\delta x$
with $\delta x=x_{u}-x_{d}$.

In practice, a series of write pulses are sent into the atomic ensembles and
the induced Stokes pulses are directed to the detectors. The time interval
between neighboring write pulses is larger than the classical communication
time. When there is a click on the detectors, the entanglement is generated
and classical information is sent back to the communication nodes to stop the
subsequent write pulses. In this case, the change of environment due to
imperfections will always induce path length fluctuations and thus phase
instability. If the entanglement between the two pairs of memory qubits is
always established at the same time, one can consider the Stokes photons
detected at the same time experience the same environment. Thus it is easy to
find $\delta x=x_{u}-x_{d}=0$ and no phase stabilization is needed.
\begin{figure}[ptb]
\begin{center}
\includegraphics[
height=5.2cm ]{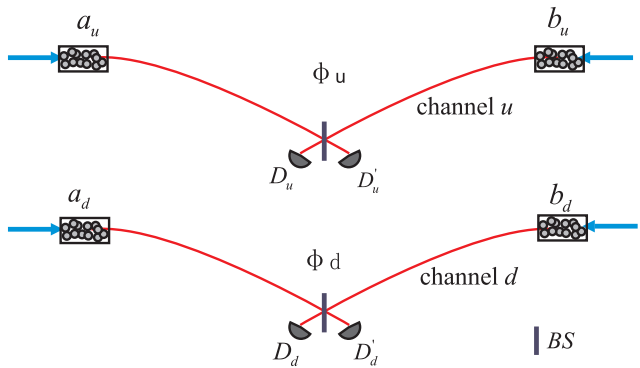}
\end{center}
\caption{In the DLCZ protocol, two entangled pairs are generated in parallel.
The relative phase between the two entangled states has to be stabilized
during the entanglement generation process.}%
\end{figure}

However entanglement generation process is probabilistic. The experiment has
to be repeated about $1/(\chi e^{-L_{0}/L_{att}} )$ times to ensure that there
is a click on the detectors. The two phases $\phi_{u}$ and $\phi_{d}%
$\ achieved at different runs of the experiments are usually
different due to the path length fluctuations in this time
interval. For instance, the entanglement between the first pair
may be constructed after the first run of the experiment, and thus
we get the phase $\phi_{u}=kx_{u}$, while the entanglement between
the second pair may be established until the last run of the
experiment, and thus we obtain the phase $\phi_{d}=kx_{d}$.
Therefore to get a high fidelity entangled pair, the relative
phase $\delta\phi=k\delta x$\ has to be stabilized during the
whole length of the communication. To stabilize the phase
instability within $\delta\phi\leq2\pi/10$, one must control the
path length instability $\delta x\leq0.1$ $\mu$m during the whole
entanglement generation process.

The path length instability is equivalent to the timing jitter of the arrival
time of the Stoke pulses after transmitting the channel over kilometer-scale
distances. To stabilize the path length instability $\delta x=c\delta
t\leq0.1$ $\mu$m, the timing jitter $\delta t$ of the Stokes pulse must be
controlled on the order of sub-femto second.

The time needed in entanglement generation process can be estimated as
follows. The distance between two communication sites is considered to be
$L_{0}=10$ km, and thus the classical communication time $T_{cc}=L_{0}/c$ is
about $33$ $\mu$s. Usually we have $2^{n}\approx100$, and thus $\chi
\approx0.0001$. In optical fibers, the photon loss rate is considered to be
$2$ dB/km for photons at a wavelength of about 800 nm, and thus the duration
$t_{0}$ of the entanglement generation process can be estimated to be about
$30$ seconds. Therefore, phase stabilization in DLCZ protocol requires that
over a timescale of about a few tens of seconds, one must control the timing
jitter after transferring a pulse sequence over several kilometers on the
order of sub-femto second. This demand is extremely difficult for current
technology. The lowest reported jitter for transferring of a timing signal
over kilometer-scale distances is a few tens of femto-seconds for averaging
times of $\geq1s$ \cite{ye}, which is 2 orders of magnitude worse than the
timing jitter needed in DLCZ protocol. In free space, the photon loss rate is
about $0.1$ dB/km and $t_{0}$ is about $0.5$ second. In this case, the path
length instability due to atmosphere fluctuations is even worse. The timing
jitter is on the order of a few nanoseconds over a timescale of $1$ second.

\subsection{Phase instability analysis II}

\begin{figure}[ptb]
\begin{center}
\includegraphics[
height=5.5cm]{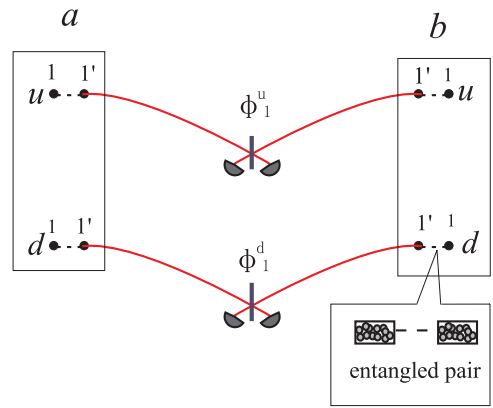}
\end{center}
\caption{Elementary entangled pairs are created locally. Entanglement swapping
is performed remotely to connect atomic ensembles between adjacent nodes $a$
and $b$.}%
\end{figure}

From the above analysis, we know that in the standard DCLZ
protocol, the requirement to stabilize the relative phase between
the two entangled pairs is severe even in the entanglement
generation stage. One may consider if entanglement generation is
performed locally, the time needed in entanglement generation
process is short and thus the requirement can be alleviated.
However, that is not the case. It is a misunderstanding that the
phase only needs to be stabilized in entanglement generation
process. In the DLCZ protocol, the single-photon Mach-Zehnder
interference is also utilized in entanglement swapping process.
When performing entanglement swapping to connect the neighboring
communication nodes, the phases have to be stabilized, too. In the
following, we will give a detailed analysis to show that the
phases between neighboring nodes have to be stabilized until the
desired remote entangled pairs are constructed.

Suppose elementary entangled pairs are created locally at each node and the
entanglement between neighboring nodes is generated via entanglement swapping,
as shown in Fig. 3. In the entanglement swapping process, one has to send
pulse sequences over a long distance and thus the path length fluctuations
have to be controlled. The two entangled pairs obtained after entanglement
swapping can be described by%
\begin{align}
|\psi_{\phi_{1}^{u}}\rangle_{a_{u_{1}},b_{u_{1}}}  & =(S_{a_{u_{1}}}^{\dagger
}+e^{i\phi_{1}^{u}}S_{b_{u_{1}}}^{\dagger})/\sqrt{2}|vac\rangle,\\
|\psi_{\phi_{1}^{d}}\rangle_{a_{d_{1}},b_{d_{1}}}  & =(S_{a_{d_{1}}}^{\dagger
}+e^{i\phi_{1}^{d}}S_{b_{d_{1}}}^{\dagger})/\sqrt{2}|vac\rangle.
\end{align}

Assume we are going to create the $up$ and $down$ entangled pairs between two
remote communication sites $A$ and $B$ at a distance of $L=2^{3}L_{0}$. The
entanglement connection process is shown step by step in Fig. 4. The entangled
pairs between neighboring nodes are created as shown in Fig. 3 and then
connected via further entanglement swapping which is also performed locally.
After 4 steps, two remote entangled pairs between sites $A$ and $B$ are
created,%
\begin{align}
|\Psi_{\Phi_{u}}\rangle_{A_{u},B_{u}}  & =(S_{A_{u}}^{\dagger}+e^{i\Phi_{u}%
}S_{B_{u}}^{\dagger})/\sqrt{2}|vac\rangle,\\
|\Psi_{\Phi_{d}}\rangle_{A_{d},B_{d}}  & =(S_{A_{d}}^{\dagger}+e^{i\Phi_{d}%
}S_{B_{d}}^{\dagger})/\sqrt{2}|vac\rangle,
\end{align}
where the accumulated phases are $\Phi_{u}=\sum_{i}\phi_{i}^{u}$ and $\Phi
_{d}=\sum_{i}\phi_{i}^{d}$. The effectively maximally entangled pair can be
described as
\begin{equation}
|\Psi_{\delta\Phi}\rangle_{PME}=(S_{A_{u}}^{\dagger}S_{B_{u}}^{\dagger
}+e^{i\delta\Phi}S_{A_{d}}^{\dagger}S_{B_{d}}^{\dagger})/\sqrt{2}|vac\rangle,
\end{equation}
where $\delta\Phi=\Phi_{u}-\Phi_{d}=\sum_{i}(\phi_{i}^{u}-\phi_{i}^{d}) $ is
the phase difference between the $up$ and $down$ entangled pairs. Note that
the phases $\phi_{i}^{u}$ or $\phi_{i}^{d}$ ($i=1,2...8$) between different
nodes are independent from each other, and thus phase stabilization requires
$\phi_{i}^{u}=\phi_{i}^{d}$ ($i=1,2...8$).

\begin{figure*}[ptb]
\begin{center}
\includegraphics[
height=10cm]{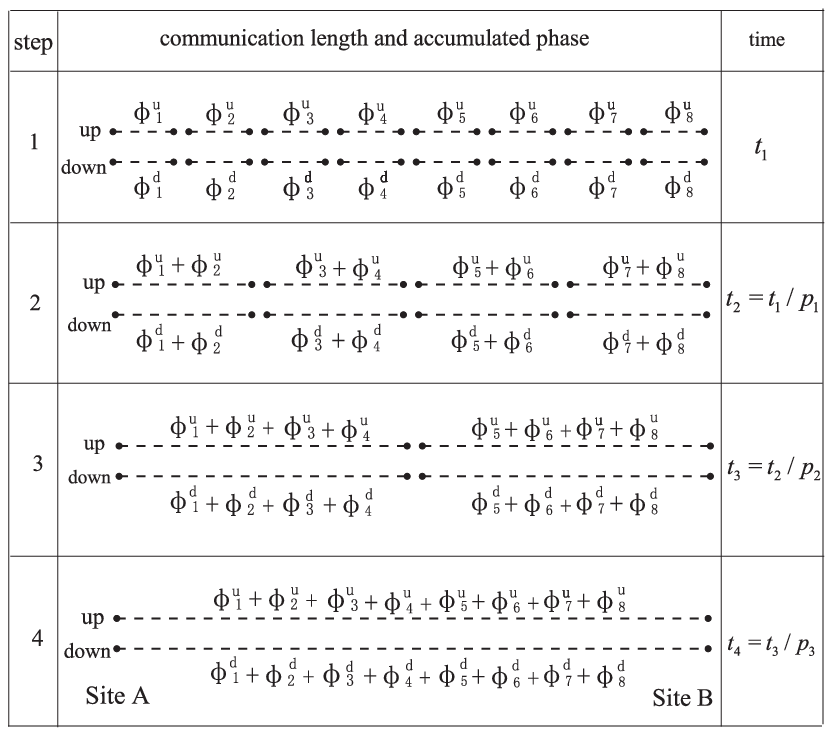}
\end{center}
\caption{ Entangled pairs are generated between neighboring communication
nodes as shown in Fig. 3. The entangled pairs are connected by performing
further entanglement swapping to construct entanglement between remote
communication sites $A$ and $B$. Entanglement connection process, as well as
the accumulated phase, is shown step by step.}%
\end{figure*}

Because entanglement swapping in every step is probabilistic, if the
entanglement swapping does not succeed in one step, one has to repeat all the
previous steps to reconstruct the entangled pairs. In this case, the phase has
to be stabilized until the desired entangled pairs $|\Psi_{\Phi_{u}}%
\rangle_{A_{u},B_{u}}$ and $|\Psi_{\Phi_{d}}\rangle_{A_{d},B_{d}}$
are both generated. For example, suppose after step 3 we have
created two $up$ entangled pairs and two $down$ entangled pairs in
parallel. In step 4, we will connect the $up$ and $down$ pairs
respectively via entanglement swapping to obtain the two desired
entangled pairs between remote sites $A$ and $B$. Since
entanglement swapping is probabilistic, it could be that we
succeed in
connecting the $up$ pairs and acquiring $|\Psi_{\Phi_{u}}\rangle_{A_{u},B_{u}%
}$, but fail to connect the $down$ pairs after performing entanglement
swapping once. In this case, we have to repeat step 1, 2 and 3 to reconstruct
the two $down$ entangled pairs and then connect them by entanglement swapping
to obtain $|\Psi_{\Phi_{d}}\rangle_{A_{d},B_{d}}$. Since the phase $\Phi_{u}$
of the $up$ pair has been fixed, the phases of the $down$ pairs $\phi_{i}^{d}$
($i=1,2...8$) have to be stabilized to satisfy $\phi_{i}^{u}=\phi_{i}^{d}$
($i=1,2...8$), until the $down$ pair $|\Psi_{\Phi_{d}}\rangle_{A_{d},B_{d}}$
is successfully generated. The total time needed in these processes is
$t_{4}=t_{1}/(p_{1}p_{2}p_{3})$. In other words, the phases $\phi_{i}^{u}$
($i=1,2...8$) and $\phi_{i}^{d}$ ($i=1,2...8$) have to be stabilized over a
time interval $t_{4}=t_{1}/(p_{1}p_{2}p_{3})$, until the desired remote
entangled pairs $|\Psi_{\Phi_{u}}\rangle_{A_{u},B_{u}}$ and $|\Psi_{\Phi_{d}%
}\rangle_{A_{d},B_{d}}$ are both generated. For long-distance
quantum communication, the total time needed is on the order of
several hours \cite{jiang}. Even in the ideal case, it is still on
the order of a few seconds. Therefore, phase stabilization in the
DLCZ protocol requires that one has to stabilize the path length
fluctuations over a long time interval after sending a pulse
sequence over kilometer-scale distances. As we discussed above, it
is extremely difficult for current technique to meet this
demanding requirement.

\section{Robust quantum repeater}

\subsection{Basic protocol}

The phase stability problem could be overcome by interfering two
photons \cite{Simon,Feng,two-photon}. Based on this, we proposed a
robust quantum repeater architecture by taking advantage of
two-photon Hong-Ou-Mandel interference \cite{zhao}. In this
section, we will give a detailed analysis on the protocol.

To exploit the advantage of two-photon interference, it is natural
to extend the DLCZ protocol by polarization encoding a memory
qubit with two atomic ensembles \cite{kuzsci,chen}, and entangling
two memory qubits at neighboring sites via a two-photon Bell-state
measurement (BSM). Unfortunately, as shown below, the BSM will not
create the desired entangled state, but a complex superposition
state with spurious contributions from second-order excitations,
which preclude further entanglement manipulation.

Let us consider two communication sites $A$ and $B$ at a distance of $L_{0}$.
A schematic setup of entanglement generation is shown in Fig. 5. Each site has
two atomic ensembles encoded as one memory qubit and the two atomic ensembles
at each node are excited simultaneously by write laser pulses. We assume the
Stokes photons generated from the two atomic ensembles at the same site have
orthogonal polarization state, e.g.,\ $|H\rangle$ and $|V\rangle$,\ which
denote horizontal and vertical linear polarization, respectively. In this way
the memory qubit is effectively entangled in the polarization states of the
emitted Stokes photons.

The Stokes photons generated from both sites are directed to the polarization
beam splitter (PBS) and subject to BSM-I at the middle point to entangle the
two neighboring memory qubits. However, the two-photon state generated in the
second-order spontaneous Raman process will also induce coincidence counts on
the detectors. Thus the BSM-I can only prepare the neighboring memory qubits
into a complex superposition state with spurious contributions from
second-order excitations. For instance, a coincidence count between D$_{1}$
and D$_{4}$ projects the two memory qubits into%
\begin{align}
|\psi\rangle_{AB}  & =[\frac{e^{i(\phi_{A}+\phi_{B})}}{2}(S_{u_{A}}^{\dagger
}S_{u_{B}}^{\dagger}+S_{d_{A}}^{\dagger}S_{d_{B}}^{\dagger})+\frac{1}%
{4}(e^{i2\phi_{A}}S_{u_{A}}^{\dagger2}\nonumber\\
& +e^{i2\phi_{B}}S_{u_{B}}^{\dagger2}-e^{i2\phi_{A}}S_{d_{A}}^{\dagger
2}-e^{i2\phi_{B}}S_{d_{B}}^{\dagger2}]|vac\rangle,
\end{align}
where $\phi_{A}$ and $\phi_{B}$ are the phases that the photons acquire,
respectively, from site $A$ and $B$ during the BSM-I. The atomic ensembles are
distinguished by subscript ($u,d$) and ($A,B$). The first part is the
maximally entangled state needed for further operations, while the second part
is the spurious two-excitation state coming from second-order excitations. The
success probability is on the order of $O(\chi^{2}\eta_{1}^{2}e^{-L_{0}%
/L_{att}})$, where $\eta_{1}$ is the detection efficiency. The time needed in
this process is $T_{0}\approx\frac{T_{cc}}{\chi^{2}\eta_{1}^{2}e^{-L_{0}%
/L_{att}}}$.

It is obvious that the phases $\phi_{A}$ and $\phi_{B}$ only lead
to a multiplicative factor $e^{i(\phi_{A}+\phi_{B})}$ before the
desired entangled state and thus have no effect on the desired
entanglement. The prize to pay is that some spurious coincidence
counts from the two-excitation terms are also registered, which
obviously prevents further entanglement manipulation and must be
eliminated by some means. However, we find that it is not
necessary to worry about these terms, because they can be
automatically washed out if the BSM in the entanglement swapping
step is carefully designed. In the ideal case a maximally
entangled state can be created by implementing entanglement
swapping.

\begin{figure}[ptb]
\begin{center}
\includegraphics[
height=5.5cm]{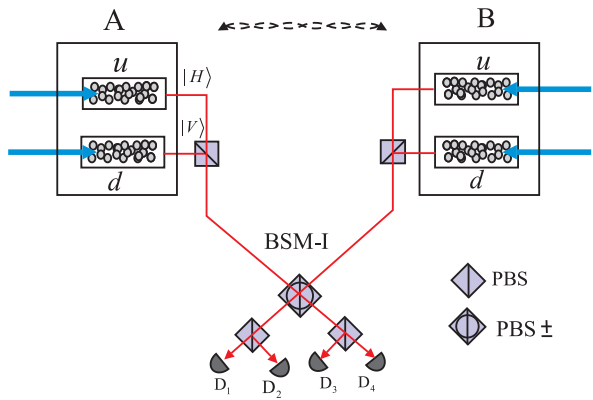}
\end{center}
\caption{Setup for entanglement generation between sites $A$ and $B$.
Forward-scattered Stokes photons, generated by an off-resonant write laser
pulse via spontaneous Raman transition, are subject to BSM-\uppercase
\expandafter{\romannumeral1} at the middle point. The Stokes photons generated
at the same site are assumed to have different polarization i.e., $|H\rangle$
and $|V\rangle$. PBS (PBS$_{\pm}$) reflects photons with polarization
$|V\rangle$ ($|-\rangle$) and transmits photons with polarization $|H\rangle$
($|+\rangle$), where $|\pm\rangle=\frac{1}{\sqrt{2}}(|H\rangle\pm|V\rangle)$.
After passing through the PBS$_{\pm}$ and PBS successively, the Stokes photons
are detected by single photon detectors. A coincidence count between single
photon detectors D$_{1}$ and D$_{4}$ (D$_{1} $ and D$_{3}$) or D$_{2}$ and
D$_{3}$ (D$_{2}$ and D$_{4} $) will project the four atomic ensembles into the
complex entangled state $|\psi\rangle_{AB}$ up to a local unitary
transformation.}%
\end{figure}

The entanglement swapping setup is depicted in Fig. 6. Let us
consider three communication sites $A,B$ and $C$, and assume that
we have created the complex entangled states (Eq. 10)
$|\psi\rangle_{AB_{L}}$ and $|\psi\rangle_{B_{R}C}$ between
($A,B_{L}$) and ($B_{R},C$), respectively \cite{note}. The memory
qubits $B_{L}$ and $B_{R}$ at site $B$ are illuminated
simultaneously by read laser pulses. The retrieved anti-Stokes
photons are subject to BSM-II. Note that the sequence of the PBSs
in BSM-II is different from BSM-I. The BSM-II is designed like
this in order that the two-photon states converted from the
spurious two-excitation terms are directed into the same output
and thus will not induce a coincidence count on the detectors. In
the ideal case, if the retrieve efficiency is unity and perfect
photon detectors are used to distinguish photon numbers, only the
two-photon coincidence count will be registered and project the
memory qubits into a maximally entangled state. For instance, when
a coincidence count between D$_{1}$ and D$_{4}$ is registered one
will obtain
\begin{equation}
|\phi^{+}\rangle_{AC}=(S_{u_{A}}^{\dagger}S_{u_{C}}^{\dagger}+S_{d_{A}%
}^{\dagger}S_{d_{C}}^{\dagger})/\sqrt{2}|vac\rangle.
\end{equation}
In this way a maximally entangled state across sites $A$ and $C$ is generated
by performing entanglement swapping. The maximally entangled state can be
extended by further entanglement swapping as usual. Both the entanglement
creation and entanglement connection in our scheme rely on two-photon
interference, so the improvement in insensitivity to path length fluctuations,
as compared to the DLCZ scheme, is about 7 orders of magnitude.

\begin{figure}[ptb]
\begin{center}
\includegraphics[
height=5.8cm]{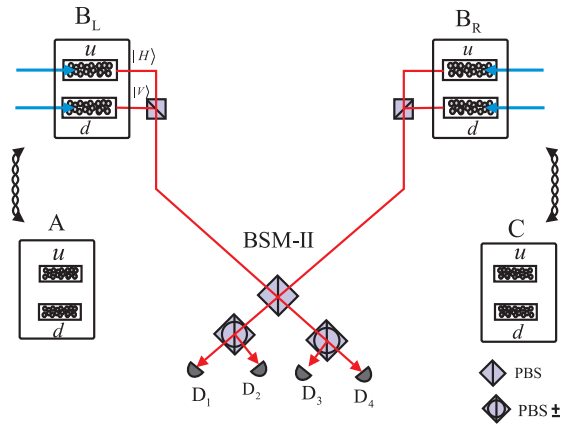}
\end{center}
\caption{Setup for entanglement connection between sites $A$ and $C$ via
entanglement swapping. Complex entangled states have been prepared in the
memory qubits between sites ($A,B_{L}$) and ($B_{R},C$). The memory qubits at
site $B$ are illuminated by near resonant read laser pulses, and the retrieved
anti-Stokes photons are subject to BSM-\uppercase
\expandafter{\romannumeral2} at the middle point. The anti-Stokes photons at
the same site have different polarizations $|H\rangle$ and $|V\rangle$. After
passing through PBS and PBS$_{\pm}$ successively, the anti-Stokes photons are
detected by single photon detectors. Coincidence counts between D$_{1}$ and
D$_{4}$ (D$_{1}$ and D$_{3}$) or D$_{2}$ and D$_{3}$ (D$_{2}$ and D$_{4}$) are
registered. The memory qubits will be projected into an effectively maximally
entangled state $\rho_{AC}$ up to a local unitary transformation.
Note that the sequence of PBSs in BSM-II is different from BSM-I.
This helps to eliminate the spurious contributions from second-order excitations.}%
\end{figure}

In practice, the retrieve efficiency $\eta_{r}$ is determined by optical depth
of the atomic ensembles \cite{optimal}, and current single photon detectors
are incapable of distinguishing photon numbers. Taking into account these
imperfections, the multi-photon coincidence counts in the BSM-II have to be
considered. Through some simple calculations, one can find that the
coincidence counts will prepare the memory qubits into a mixed entangled state
of the form%
\begin{equation}
\rho_{AC}=p_{2}\rho_{2}+p_{1}\rho_{1}+p_{0}\rho_{0},
\end{equation}
where the coefficients $p_{2}$, $p_{1}$ and $p_{0}$ are determined by the
retrieve efficiency and detection efficiency (see Appendix 1 for details).
Here $\rho_{2}=|\phi^{+}\rangle_{AC}\langle\phi^{+}|$ is a maximally entangled
state, $\rho_{1}$ is a maximally mixed state, where only one of the four
atomic ensembles has one excitation, and $\rho_{0}$ is the vacuum state that
all the atomic ensembles are in the ground states.

It is easy to see that $\rho_{AC}$ is in fact an effectively
maximally entangled states, which can be projected automatically
to a maximally entangled state in the entanglement-based quantum
cryptography schemes. When implementing quantum cryptography via
the Ekert protocol \cite{Ekert}, we randomly choose the detection
basis at the remote sites and detect the photons retrieved from
the atomic ensembles. Then we compare the detection basis by
classical communication. In this process, only the coincidence
counts are registered and used for quantum cryptography. In our
case only the first term $\rho_{2}$ will contribute to a
coincidence count between the detectors at the two sites and will
be registered after classical communication. The maximally mixed
state term $\rho_{1}$ and the vacuum term $\rho_{0}$ have no
contribution to the experimental results, and thus
$\rho_{AC}$ is equivalent to the Bell state $|\phi^{+}\rangle_{AC}=(S_{u_{A}%
}^{\dagger}S_{u_{C}}^{\dagger}+S_{d_{A}}^{\dagger}S_{d_{C}}^{\dagger}%
)/\sqrt{2}|vac\rangle$.

\subsection{Entanglement connection and scalability}

The effectively entangled state can be connected to longer communication
distance via further entanglement swapping. To implement a quantum repeater
protocol, a nesting scheme is used in entanglement connection process
\cite{repeater1,repeater2}. Taking into account higher-order excitations, the
effectively mixed entangled pair reads $\rho^{\prime}=\rho+p_{2}^{\prime}%
\rho_{2}^{\prime}+p_{3}^{\prime}\rho_{3}^{\prime}$, where the normalized
density matrix $\rho_{2}^{\prime}$ and $\rho_{3}^{\prime}$ denote the
two-excitation mixed state and three-excitation mixed state generated due to
higher-order excitations in the spontaneous Raman process, and the small
coefficients $p_{2}^{\prime}$ and $p_{3}^{\prime}$ are on the order of
$O(\chi)\ll1$. After the $j$-th swapping step, the effectively entangled pair
can be described as (see Appendix 2)
\begin{equation}
\rho_{s_{j}}^{\prime}=p_{2s_{j}}\rho_{2s_{j}}+p_{1s_{j}}\rho_{1s_{j}%
}+p_{0s_{j}}\rho_{0s_{j}}+p_{2s_{j}}^{\prime}\rho_{2s_{j}}^{\prime}+p_{3s_{j}%
}^{\prime}\rho_{3s_{j}}^{\prime}.
\end{equation}
Here $\rho_{2s_{j}}$ is the maximally entangled state between two memory
qubits at a distance of $L=2^{j}L_{0}$, and $\rho_{1s_{j}}$ and $\rho_{0s_{j}%
}$ are also the maximally mixed state and vacuum state, respectively. Note
that $\rho_{s_{1}}^{\prime}=\rho^{\prime}$ is just the mixed entangled state
created after the first entanglement swapping step. The coefficients can be
estimated to be
\begin{align}
p_{2s_{j}}^{\prime}  & \sim O(j\chi),\text{ }p_{3s_{j}}^{\prime}\sim
O(\chi),\\
p_{\alpha s_{j}}  & \approx p_{\alpha s_{j-1}}+O(j\chi),\ \ (\alpha=0,1,2).
\end{align}
From Eq. (14), it is easy to see that the contributions from
higher-order excitations $\rho_{2s_{j}}^{\prime}$ and
$\rho_{3s_{j}}^{\prime}$ can be safely neglected, as long as the
small excitation probability fulfills $j\chi\ll1$, which can be
easily achieved by tuning the write laser pulse. One can also see
that the coefficients $p_{2s_{j}}$, $p_{1s_{j}}$ and $p_{0s_{j}}$
are stable to the first order, therefore the probability to find
an entangled pair in the remaining memory qubits is almost a
constant and will not decrease significantly with distance during
the entanglement connection process. The time needed for the
$j$-th connection step satisfies the iteration formula
$T_{s_{j}}=\frac{1}{p_{s_{j}}}[T_{s_{j-1}}+2^{j-1}T_{cc}]$ with
$p_{s_{j}}$ the success probability of the \textit{j-}th swapping
step. The total time needed for the entanglement connection
process is
\begin{equation}
T_{tot}\approx T_{0}\prod\limits_{j}p_{s_{j}}^{-1}\approx\frac{T_{cc}}%
{\chi^{2}\eta_{1}^{2}}e^{L_{0}/Latt}(L/L_{0})^{\log_{2}^{1/\eta}},
\end{equation}
where $\eta=\eta_{r}^{2}\eta_{1}^{2}$ is a constant. The excitation
probability can be estimated to be $\chi\sim L_{0}/L$, and then the time
needed in the entanglement connection process $T_{tot}\propto$ $(L/L_{0}%
)^{2+\log_{2}^{1/\eta}}$ scales polynomially or quadratically with the
communication distance.

One can modify our protocol by performing entanglement generation
locally and entanglement swapping remotely. It will help to
increase the scalability, since entanglement generation is usually
the rate-limiting stage due to the low excitation probability.
Local entanglement can also be generated via the standard DLCZ
protocol and then connected by two-photon Hong-Ou-Mandel
interference, because local path length fluctuations can be well
controlled. The experimental setup is shown in Fig. 7. Here we
need BSM-I to eliminate spurious contributions from high-order
excitations. Note that the setup in Fig. 7 is a simple variation
of the scheme due to Jiang \textit{et al}. \cite{jiang}, where
entanglement generation is performed remotely and entanglement
swapping is performed locally. We remark that this simple
modification is crucial to long-distance quantum communication, as
entanglement generation relies on single-photon interference and
must be performed locally.

\subsection{Alternative approach}

\begin{figure}[ptb]
\begin{center}
\includegraphics[
height=5.5cm]{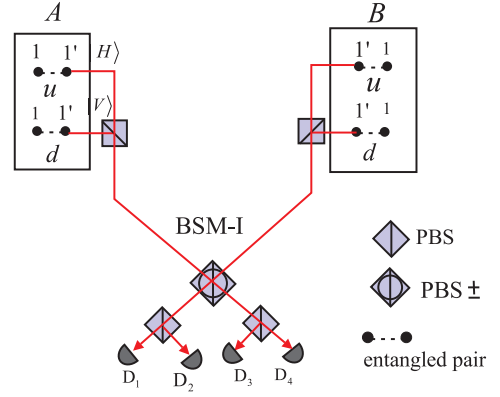}
\end{center}
\caption{Elementary entangled pairs are first locally generated
via the standard DLCZ protocol. The anti-Stokes photons are
subject to BSM-\uppercase \expandafter{\romannumeral1} to connect
neighboring communication nodes. We also assume the anti-Stokes
photons retrieved from atomic ensembles at the same site have
different polarization. Note that BSM-\uppercase
\expandafter{\romannumeral1} also helps to eliminate the spurious
contributions from higher order excitations.}%
\end{figure}

The locally entangled memory qubits can be generated by other
means. Atomic ensembles can also serve as a quantum memory to
store a photonic state \cite{lukin,Fleis}. By applying a time
dependent classical control laser pulse of a Rabi frequency
$\Omega_{c}$, the whole system has a particular zero-energy
eigenstate, i.e., the dark-state-polariton. The single-polariton
state is $|D,1\rangle=\frac{\Omega_{c}(t)}{\sqrt{\Omega
_{c}^{2}(t)+g^{2}N}}|1\rangle_{p}|0\rangle_{a}-\frac{g\sqrt{N}}{\sqrt
{\Omega_{c}^{2}(t)+g^{2}N}}|0\rangle_{p}S^{\dagger}|0\rangle_{a}$,
with $g$ being the coupling constant for the $|g\rangle-|e\rangle$
transition. Here $\left\vert 0\right\rangle _{p}$ ($\left\vert
1\right\rangle _{p}$) is the vacuum (single-photon) state of the
quantized field to be stored. The quantum memory works by
adiabatically changing $\Omega_{c}(t)$ such that one can
coherently map $|D,1\rangle$ onto either purely atom-like state
$|0\rangle _{p}S^{\dagger}|0\rangle_{a}$ where the single photon
is stored, or purely photon-like state
$|1\rangle_{p}|0\rangle_{a}$, which corresponds to the release of
the single photon.

\begin{figure}[ptb]
\begin{center}
\includegraphics[
height=5cm]{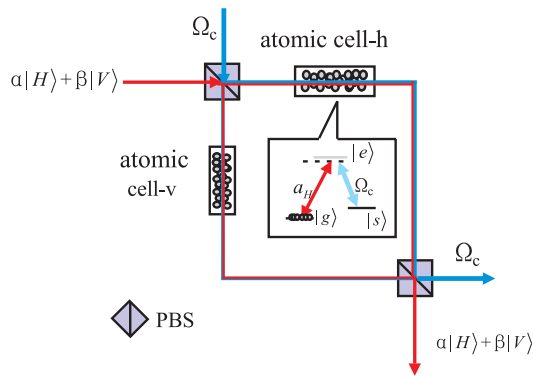}
\end{center}
\caption{Quantum memory for photonic polarization qubits. Two ensembles are
driven by a classical control field. Classical and quantized light fields are
fed into the first PBS and will leave at two different outputs of the second
PBS. As each atomic cell works as quantum memory for single photons with
polarization $|H\rangle$ or $|V\rangle$ via the adiabatic transfer method, the
whole setup is then quantum memory of any single-photon polarization states.
The inset shows the relevant level structure of the atoms. The $|e\rangle
-|s\rangle$ transition is coherently driven by the classical control field of
Rabi frequency $\Omega_{c}$, and the $|g\rangle-|e\rangle$ transition is
coupled to a quantized light field.}%
\end{figure}

To exploit the advantage of two-photon Hong-Ou-Mandel-type
interference, we need a quantum memory for the photonic
polarization qubits. Figure 8 shows quantum memory for storing any
single-photon polarization states by the dark-state-polariton
method. Two atomic ensembles being a quantum memory for
polarization qubits at each node are thus the required localized
memory qubit in our scheme. Thus transformation between an
arbitrary photon polarization state $\alpha\left\vert
H\right\rangle +\beta\left\vert V\right\rangle $ and the
corresponding state stored in atomic ensembles $(\alpha
S_{h}^{\dagger }+\beta S_{v}^{\dagger})\left\vert 0\right\rangle $
can be achieved by adiabatically manipulating the control laser
pulse. Importantly, our quantum memory works even when the two
probability amplitudes in the stored state $\alpha\left\vert
H\right\rangle +\beta\left\vert V\right\rangle $\ are not
c-numbers but quantum states of other photonic qubits. As a
result, two memory qubits $U$ and $D$ at one site (see Fig. 9a)
can be deterministically entangled in their \textquotedblleft
polarizations\textquotedblright\ by
storing two polarization-entangled photons, e.g., $\frac{1}{\sqrt{2}}%
(S_{h_{U}}^{\dagger}S_{h_{D}}^{\dagger}+S_{v_{U}}^{\dagger}S_{v_{D}}^{\dagger
})|vac\rangle\leftrightarrow\frac{1}{\sqrt{2}}(\left\vert H\right\rangle
\left\vert H\right\rangle +\left\vert V\right\rangle \left\vert V\right\rangle
)$. The latter are entangled by a deterministic polarization-entangler using
four single photons, linear optics and an event-ready detection. With an
overall success probability of $\frac{1}{8}$ for perfect photon counting, such
an \textquotedblleft event-ready\textquotedblright\ entangler can
deterministically generate two maximally polarization-entangled qubits (see
Appendix 3).

\begin{figure}[ptb]
\begin{center}
\includegraphics[
height=7cm]{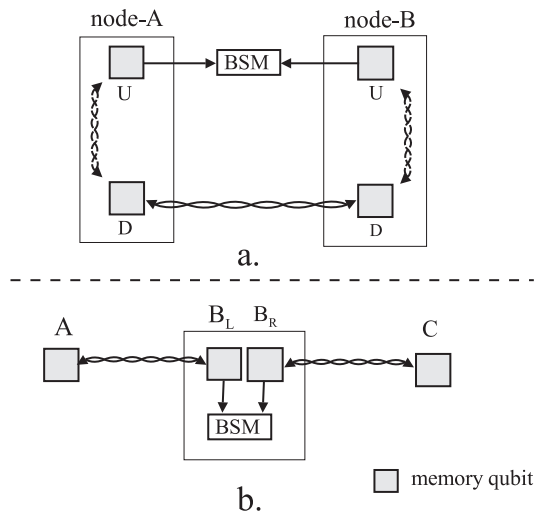}
\end{center}
\caption{a). Entanglement swapping between adjacent communications
nodes $A$ and $B$. Two pairs of entangled memory qubits are first
generated by storing the event-ready entanglement of two photons
at each node. Then the two photons stored in the $U$ memory at the
two nodes are simultaneously retrieved and subject to a two-photon
BSM at the middle point. This entanglement swapping process will
in an event-ready way entangle the two distant $D$ memory qubits.
b). Entanglement connection to extend the communication length.
Two well entangled pairs of memory qubits, one across nodes $(A,
B_{L})$ and $(B_{R},C)$ are prepared in parallel. The BSM on the
two photons released simultaneously from the two memories at node
$B$ results in, with a probability of 1/2 , well entangled
quantum memories across nodes $A$ and $C$ in a definite Bell state.}%
\end{figure}

Polarization encoding allows a two-photon interference
entanglement swapping to construct entanglement between adjacent
sites. As shown in Fig. 9, one can first create each memory pair
in maximal event-ready entanglement at two adjacent communication
nodes and then the two photons stored in the two $U$ memories are
simultaneously retrieved and subject to a two-photon BSM at the
middle point. Conditioned on the result of this BSM, the remaining
two $D$ memory qubits are maximally entangled, also in an
event-ready way. Usual entanglement swapping can be applied to the
polarization encoding memory qubits and thus allows the
implementation of a robust quantum repeater.

\subsection{Entanglement purification}

With imperfect entanglement and erroneous local operations, entanglement
connection, together with decoherence, will reduce the fidelity of
entanglement. Then at certain stage of entanglement connection, the less
entangled states have to be purified via the entanglement purification
protocol \cite{purify1,purify2} to enable further entanglement connection.
Fig. 10 shows how to achieve linear optical entanglement purification between
any specified two nodes, e.g., node-$I$ and node-$J$, across which one has
less entangled pairs of quantum memories.

Assume two effectively mixed entangled pairs of fidelity $F$ are created in
parallel via entanglement connection as we discussed above. The effectively
entangled states stored in the four memory qubits are converted into entangled
photons by the read laser pulses, and then subject to two PBSs, respectively.
The photons in mode $b_{1}$ and $b_{2}$ are detected in $|\pm\rangle=\frac
{1}{\sqrt{2}}(|H\rangle\pm|V\rangle)$ basis by single photon detectors, and
will project the photons in mode $a_{1}$ and $a_{2}$ into an effectively
maximally entangled state of higher fidelity $F^{\prime}$ (see Appendix 4).
The higher-fidelity entangled pair in mode $a_{1}$ and $a_{2}$ can be restored
into two distant memory qubits at nodes $I$ and $J$ by means of the
dark-state-polariton method for further manipulation. \begin{figure}[ptb]
\begin{center}
\includegraphics[
height=5cm]{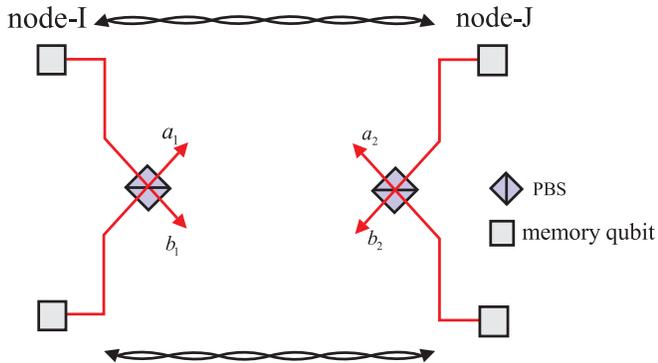}
\end{center}
\caption{Setup for quantum entanglement purification. Entangled states have
been prepared in the memory qubits between two distant nodes $I$ and $J$. The
memory qubits at the two sites are illuminated by near resonant read laser
pulse, and the retrieved entangled photon pairs are directed to two PBS
respectively. The photons in mode $b_{1}$ and $b_{2}$ are detected in
$|\pm\rangle=\frac{1}{\sqrt{2}}(|H\rangle\pm|V\rangle)$ basis and the
remaining photons in mode $a_{1}$ and $a_{2}$ are restored in the memory
qubits at the two sites respectively.}%
\end{figure}

To generate a remote entangled pair, the nested quantum purification has to be
implemented \cite{repeater1,repeater2}. The total time overhead to create
entanglement across two communication nodes at a distance of 1280 km can be
numerically estimated. In our calculation, we assume the distance $L_{0}=10$
km and the photon loss rate is $0.1$ dB/km in free space. To improve the
scalability, we assume entanglement generation is performed locally and the
entanglement generation time is considered to be $100$ $\mu s$. The fidelity
of the adjacent entangled memory qubits is $F = 0.88$, as can be estimated by
connecting two adjacent memories from two pairs of photon-memory entanglement
after $5$ km free space transmission of both photons \cite{peng}. One of the
major factors affecting the efficiency of our scheme is single-photon
detection. Fortunately, high-efficiency photon counting is feasible by using
quantum state transfer and state-selective fluorescence detection with nearly
unit efficiency \cite{kwait,imm}. To increase the efficiency, we assume photon
counting detectors with detection efficiency $99\%$ are used, and the retrieve
efficiency is considered to be $98\%$. Entanglement purification is performed
three times during the entanglement connection process to improve the
fidelity. Our numerical results give a total time of about 23 seconds to
create an effectively entangled pair, with a probability of $0.75$ to get the
entangled pair of fidelity $94\%$.

\subsection{Comparison between different schemes }

Recently, several atomic ensemble based quantum repeater schemes
were proposed building on the DLCZ protocol. These schemes still
have phase stability problem since single-photon interference is
also used in some stages. The scheme presented in Ref.
\cite{jiang} is similar to our protocol, where they gave a
detailed analysis on the superior scalability of polarization
encoding. However, single-photon interference is used in
entanglement generation process, and thus entanglement generation
should be performed locally. As we have discussed in
Sec.\uppercase\expandafter{\romannumeral3}, a simple modification
of their original scheme allows the implementation of a robust
quantum repeater. In Ref. \cite{simon2}, Simon \textit{et al.}
proposed a quantum repeater, where they suggested to make
entanglement generation attempts many times with the help of
photon pairs and multi-mode memories. The use of multi-mode
memories promises a speedup in entanglement generation by several
orders of magnitude. However, entanglement generation and
entanglement swapping in this protocol need single-photon
interference. The phase stabilization problem can be overcome by
using those cases where the entanglement swapping succeeds at the
same time for the upper and lower chains \cite{simon3}. Besides,
the fidelity of the final entanglement is sensitive to phase
instability due to the lack of entanglement purification. It was
pointed out that in this type of protocol, an initial small phase
error will induce the final entanglement fidelity no more than
65\% \cite{jiang}.

The ideas of polarization encoding, two-photon BSM and active
entanglement purification presented in our protocol is crucial to
long distance quantum communication. The combination of these
ideas enables a realistic fault-tolerant quantum repeater with
atomic ensembles and linear optics.

\section{Conclusion}

In summary, we have given a particular analysis on phase stability problem of
the DLCZ protocol. This problem can be overcome by taking advantage of
two-photon Hong-Ou-Mandel-type interference, which alleviates the phase
stability requirements by about 7 orders of magnitude. Most of the ingredients
in our protocol have been experimentally realized in recent years
\cite{yuan,chen}. A long storage time is crucial for implementing atomic
ensemble based quantum repeater protocol. Storage time of up to $30$ $\mu s$
was reported recently \cite{kuzmich}. An optical dipole trap may have the
potential to extend the storage time to $1$ second. According to a recent
proposal, quantum memory with nuclear atomic spins might have very long
storage time of about hours \cite{hours}. Our scheme also relies on the
ability to reliably transfer of photon's polarization states over a free-space
or optical fiber channel. Two recent experiments demonstrated this ability up
to 100 km in free space \cite{free} and in fiber \cite{peng2}. Our scheme
faithfully implements a robust quantum repeater and thus enables a realistic
avenue for relevant long-distance quantum communication.

\begin{acknowledgments}
We are grateful to Q. Zhang and X.-M. Jin for helpful discussions. This work
was supported by the National Fundamental Research Program (under grant No.
2006CB921900), the CAS, the NNSFC, the DFG, the Alexander von Humboldt
Foundation, and the European Commission. Y.-A. Chen acknowledges additional
support from the Deutsche Telekom Stiftung.
\end{acknowledgments}

\appendix*

\section{Evaluation of the coefficients}

\subsection{Entanglement swapping}

In practice, the retrieve efficiency $\eta_{r}$ is limited by the optical
depth of the atomic ensemble and single photon detectors can't resolve photon
number. Thus the three-photon and four-photon coincidences are also registered
when performing BSM-\uppercase\expandafter{\romannumeral2}, which will result
in an effectively entangled state $\rho=p_{2}\rho_{2}+p_{1}\rho_{1}+p_{0}%
\rho_{0}$. The unnormalized coefficients are calculated to be
\begin{align}
p_{2}^{(u)}  & =\frac{\eta_{r}^{2}\eta_{1}^{2}}{32},\\
p_{1}^{(u)}  & =\frac{\eta_{r}^{2}(1-\eta_{r})\eta_{1}^{2}}{16}+\frac{\eta
_{r}^{3}}{32}(\frac{\eta_{1}\eta_{2}}{2}+\eta_{1}^{2}),\\
p_{0}^{(u)}  & =\frac{\eta_{r}^{3}}{32}(1-\eta_{r})(\frac{1}{2}\eta_{1}%
\eta_{2}+\eta_{1}^{2})+\nonumber\\
& \frac{\eta_{r}^{2}(1-\eta_{r})^{2}\eta_{1}^{2}}{32}+\frac{\eta_{r}^{4}}%
{64}(\frac{1}{4}\eta_{2}^{2}+\eta_{1}^{2}),
\end{align}
where $\eta_{1}$ and $\eta_{2}$ are the detector efficiency for single photon
state and two photon state. The success probability of entanglement swapping
is $p=p_{2}^{(u)}+p_{1}^{(u)}+p_{0}^{(u)}$.

\subsection{Entanglement connection}

Considering high-order excitations in the spontaneous Raman process, the
effectively entangled pair can be described by $\rho^{\prime}=\rho
+p_{2}^{\prime}\rho_{2}^{\prime}+p_{3}^{\prime}\rho_{3}^{\prime}$. Here we
introduce two-excitation density matrix $\rho_{2}^{\prime}$, containing the
terms $S_{u_{A}}^{\dagger2},S_{d_{C}}^{\dagger2},S_{u_{A}}^{\dagger}S_{d_{A}%
}^{\dagger},S_{u_{A}}^{\dagger}S_{u_{C}}^{\dagger}$ \textit{etc.}, and
three-excitation density matrix $\rho_{3}^{\prime}$, containing the terms
$S_{u_{A}}^{\dagger2}S_{u_{C}}^{\dagger}$, $S_{d_{C}}^{\dagger2}S_{u_{A}%
}^{\dagger}$, $S_{u_{A}}^{\dagger}S_{u_{C}}^{\dagger}S_{d_{C}}^{\dagger}$,
$S_{u_{A}}^{\dagger}S_{d_{A}}^{\dagger}S_{u_{C}}^{\dagger}$ \textit{etc.}, to
denote the contributions from higher-order excitations. The small efficient
$p_{2}^{\prime}$ and $p_{3}^{\prime}$ are on the order of $O(\chi)\ll1$. When
implementing nesting entanglement connection, the effectively entangled state
reads $\rho_{s_{j}}^{\prime}=p_{2s_{j}}\rho_{2s_{j}}+p_{1s_{j}}\rho_{1s_{j}%
}+p_{0s_{j}}\rho_{0s_{j}}+p_{2s_{j}}^{\prime}\rho_{2s_{j}}^{\prime}+p_{3s_{j}%
}^{\prime}\rho_{3s_{j}}^{\prime}$. The unnormalized coefficients can be
calculated to be
\begin{align}
p_{2s_{j}}^{(u)}  & \approx\frac{1}{2}p_{2s_{j-1}}^{2}\eta,\\
p_{1s_{j}}^{(u)}  & \approx\frac{1}{2}\eta\lbrack p_{1s_{j-1}}p_{2s_{j-1}%
}+O(p_{2s_{j-1}}p_{2s_{j-1}}^{\prime})\nonumber\\
& +O(p_{1s_{j-1}}p_{2s_{j-1}}^{\prime})+O(p_{0s_{j-1}}p_{3s_{j-1}}^{\prime
})],\\
p_{0s_{j}}^{(u)}  & \approx\frac{1}{8}\eta\lbrack p_{1s_{j-1}}^{2}%
+O(p_{0s_{j-1}}p_{2s_{j-1}}^{\prime})],\\
p_{2s_{j}}^{\prime(u)}  & \sim O(p_{2s_{j-1}}p_{2s_{j-1}}^{\prime2}%
\eta)+O(p_{1s_{j-1}}p_{3s_{j-1}}^{\prime}\eta),\\
p_{3s_{j}}^{\prime(u)}  & \sim O(p_{2s_{j-1}}p_{3s_{j-1}}^{\prime}\eta),
\end{align}
with $\eta=\eta_{1}^{2}\eta_{r}^{2}$, where the three-photon coincidence
counts are safely neglected. From the above equations, we find that%
\begin{align}
\frac{p_{3s_{j}}^{\prime(u)}}{p_{2s_{j}}^{(u)}}  & \sim O(\frac{p_{3s_{j-1}%
}^{\prime}}{p_{2s_{j-1}}})\sim O(\frac{\chi}{p_{2}}),\\
\frac{p_{2s_{j}}^{\prime(u)}}{p_{2s_{j}}^{(u)}}  & \sim O(\frac{p_{2s_{j-1}%
}^{\prime}}{p_{2s_{j-1}}})+O(\frac{p_{3s_{j-1}}^{\prime}}{p_{2s_{j-1}}})\sim
O(\frac{j\chi}{p_{2}}),\\
\frac{p_{1s_{j}}^{(u)}}{p_{2s_{j}}^{(u)}}  & \approx\frac{p_{1s_{j-1}}%
}{p_{2s_{j-1}}}+O(\frac{p_{3s_{j-1}}^{\prime}}{p_{2s_{j-1}}})+O(\frac
{p_{2s_{j-1}}^{\prime}}{p_{2s_{j-1}}}),\\
\frac{p_{0s_{j}}^{(u)}}{p_{2s_{j}}^{(u)}}  & \approx\frac{1}{4}(\frac
{p_{1s_{j-1}}}{p_{2s_{j-1}}})^{2}+O(\frac{p_{2s_{j-1}}^{\prime}}{p_{2s_{j-1}}%
}),
\end{align}
where we have considered the coefficients $p_{2s_{j-1}}$, $p_{1s_{j-1}}$, and
$p_{0s_{j-1}}$ are on the same order of magnitude. Finally, we conclude that
during the nesting entanglement connection process, the coefficients can be
estimated to be%
\begin{align}
p_{3s_{j}}^{\prime}  & \sim O(\chi),p_{2s_{j}}^{\prime}\sim O(j\chi),\\
p_{\alpha s_{j}}  & \approx p_{\alpha s_{j-1}}+O(j\chi).
\end{align}
The success probability of the $j$-th entanglement connection is $p_{s_{j}%
}=p_{2s_{j}}^{(u)}+p_{1s_{j}}^{(u)}+p_{0s_{j}}^{(u)}$.

\subsection{Deterministic entangler}

\begin{figure}[ptb]
\begin{center}
\includegraphics[
height=5cm]{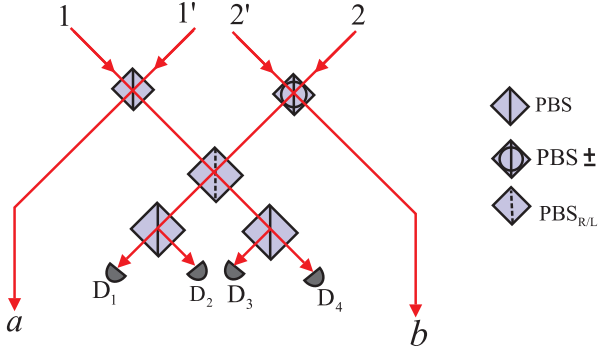}
\end{center}
\caption{Deterministic single-photon polarization entangler. PBS (PBS$_{\pm}$;
PBS$_{R/L}$) reflects photons with vertical polarization $|V\rangle
$($|-\rangle$; $|L\rangle$ and transmits photons with horizontal-polarization
$|H\rangle$ ($|+\rangle$;$|R\rangle$). Here $|\pm\rangle=\frac{1}{\sqrt{2}%
}(|H\rangle+|V\rangle)$;$|R/L\rangle=\frac{1}{\sqrt{2}}(|H\rangle\pm
i|V\rangle)$. The four single photons are prepared on demand in an initial
state $|-\rangle_{1}|V\rangle_{2}|+\rangle_{1^{\prime}}|H\rangle_{2^{\prime}}%
$. After passing through the first PBS and PBS$_{\pm}$, one selects the
`four-mode' case where there is one and only one photon in each of the four
output modes. Then the BSM will collapse photons in modes $a $ and $b$ into a
Bell state conditioned on the result of the BSM. In our case, a coincidence
count between single-photon detectors D1 and D4 (D1 and D3) or between D2 and
D3 (D2 and D4) leaving photons along paths $a$ and $b$ deterministically
entangled in $|\psi^{+}\rangle_{ab}$($|\phi^{-}\rangle_{ab}$).}%
\end{figure}

The deterministic single-photon polarization entangler is depicted
in Fig. 11. The
input state is $|-\rangle_{1}|V\rangle_{2}|+\rangle_{1^{\prime}}%
|H\rangle_{2^{\prime}}$. In the ideal case where single photons
can be created on demand and photon-number counting detectors are
used to identify the Bell
states, we will obtain two maximally entangled photons in $|\psi^{+}%
\rangle_{ab}$ or $|\phi^{-}\rangle_{ab}$, conditioned on a coincidence count
in two of the four detectors with a success probability $\frac{1}{8}$
\cite{event}.

However, current single photon sources are probabilistic and the mostly used
single photon detectors cannot distinguish between one and more than one
detected photons. Due to these imperfections, the output state in $a$ and $b$
is not a pure state but a mixed entangled state. Assuming the single photon
sources can generate single photons with probability $p_{r}$, it is easy to
see that when there are $2$ photons ($\{1,2\},\{1,2^{\prime}\},\{1^{\prime
},2\},\{1^{\prime},2^{\prime}\}$) with probability $p_{r}^{2}(1-p_{r})^{2}$,
$3$ photons ($\{1,1^{\prime},2\},\{1,1^{\prime},2^{\prime}\},\{1,2,2^{\prime
}\},\{1^{\prime},2,2^{\prime}\}$) with probability $p_{r}^{3}(1-p_{r})$ and
$4$ photons ($\{1,1^{\prime},2^{\prime},2^{\prime} $\}) with probability
$p_{r}^{4}$ emitted from single photon sources, there will be a coincidence
count between two of the detectors.

Considering all these possibilities, we find that if one of the four
coincidence counts occurs, e.g., D$_{1}$ and D$_{4}$ is registered, the output
state in $a$ and $b$ is equivalent to an effectively maximally entangled
state
\begin{equation}
\rho_{c}=p_{2c}\rho_{2c}+p_{1c}\rho_{1c}+p_{0c}\rho_{0c},
\end{equation}
with the unnormalized coefficients
\begin{align}
p_{2c}^{(u)} &  =\frac{p_{r}^{4}\eta_{1}^{2}}{32},\\
p_{1c}^{(u)} &  =\frac{p_{r}^{3}(1-p_{r})\eta_{1}^{2}}{8}+\frac{p_{r}^{4}%
\eta_{1}^{2}}{32}+\frac{p_{r}^{4}\eta_{1}\eta_{2}}{64},\\
p_{0c}^{(u)} &  =\frac{1}{32}[p_{r}^{3}(1-p_{r})(2\eta_{1}^{2}+\eta_{1}%
\eta_{2})\nonumber\\
&  +p_{r}^{4}\eta_{1}\eta_{2}+4p_{r}^{2}(1-p_{r}^{2})\eta_{1}^{2}].
\end{align}
Here $\rho_{2c}$ is one of the maximally entangled Bell states, $\rho_{1c}$ is
the one-photon maximally mixed state and $\rho_{0c}$ is the vacuum state,
which indicates that all the input photons are detected and there is no photon
in the output $a$ and $b$. After the event-ready mixed entangled state is
successfully generated, it will be directed and stored into memory qubits at
each communication node as discussed in the main text. The success probability
for the event-ready entangler is $p_{c}=p_{2c}^{(u)}+p_{1c}^{(u)}+p_{0c}%
^{(u)}$.

\subsection{Entanglement purification}

Suppose we have generated an effectively mixed entangled state $\rho
_{m}=p_{2m}\rho_{2m}+p_{1m}\rho_{1m}+p_{0m}\rho_{0m}$ of fidelity $F$ across
nodes $I$ and $J$. For simplicity, we assume the mixed state is of the form
$\rho_{2m}=F|\phi^{+}\rangle_{IJ}\langle\phi^{+}|+(1-F)|\psi^{+}\rangle
_{IJ}\langle\psi^{+}|$, with $|\phi^{+}\rangle_{IJ}=(S_{u_{I}}^{\dagger
}S_{u_{J}}^{\dagger}+S_{d_{I}}^{\dagger}S_{d_{J}}^{\dagger})/\sqrt
{2}|vac\rangle$ and $|\psi^{+}\rangle_{ij}=(S_{u_{I}}^{\dagger}S_{d_{J}%
}^{\dagger}+S_{d_{I}}^{\dagger}S_{u_{J}}^{\dagger})/\sqrt{2}|vac\rangle$. Two
pairs of entangled memory qubits are generated in parallel. Linear optical
entanglement purification will project the photons in mode $a_{1}$ and $a_{2}$
into a mixed entangled state of higher fidelity $F^{\prime}=\frac{F^{2}}%
{F^{2}+(1-F)^{2}}$, which can be described as%

\begin{equation}
\rho_{p}=p_{2p}\rho_{2p}+p_{1p}\rho_{1p}+p_{0p}\rho_{0p},
\end{equation}
with $\rho_{2p}=F^{\prime}|\phi^{+}\rangle_{IJ}\langle\phi^{+}|+(1-F^{\prime
})|\psi^{+}\rangle_{IJ}\langle\psi^{+}|$. The unnormalized coefficients are
\begin{align}
p_{2p}^{(u)} &  =\frac{1}{2}p_{2m}^{2}\eta_{r}^{4}\eta_{1}^{2}[F^{2}%
+(1-F)^{2}],\\
p_{1p}^{(u)} &  =p_{2m}^{2}\eta_{r}^{3}(1-\eta_{r})\eta_{1}^{2}+\frac{1}%
{2}p_{1m}p_{2m}\eta_{r}^{3}\eta_{1}^{2}\nonumber\\
&  +p_{2m}^{2}\eta_{r}^{4}F(1-F)\eta_{1}\eta_{2},\\
p_{0p}^{(u)} &  =p_{2m}^{2}[\frac{1}{4}\eta_{r}^{4}F^{2}\eta_{2}^{2}+\eta
_{r}^{3}(1-\eta_{r})F(1-F)\eta_{1}\eta_{2}\nonumber\\
&  +\eta_{r}^{2}(1-\eta_{r})^{2}(F+1/2)\eta_{1}^{2}+\eta_{r}^{3}(1-\eta
_{r})F^{2}\eta_{1}\eta_{2}]\nonumber\\
&  +p_{2m}p_{1m}[\eta_{r}^{2}(1-\eta_{r})(F_{m}+1/2)\eta_{1}^{2}+\eta_{r}%
^{3}\frac{F}{2}\eta_{1}\eta_{2}]\nonumber\\
&  +\frac{1}{8}p_{1m}^{2}\eta_{r}^{2}\eta_{1}^{2}+p_{2m}p_{0m}\eta_{r}%
^{2}F\eta_{1}^{2}.
\end{align}
The success probability of entanglement purification is $p_{p}=p_{2p}%
^{(u)}+p_{1p}^{(u)}+p_{0p}^{(u)}$.


\begin{thebibliography}{99}
\bibitem {Gisin}N. Gisin \textit{et al.}, Rev. Mod. Phys. \textbf{74}, 145 (2002).

\bibitem {Ekert}A.K. Ekert, Phys. Rev. Lett. \textbf{67}, 661 (1991).

\bibitem {marek} M. Zukowski, A. Zeilinger, M. A. Horne, and A.
Ekert, Phys. Rev. Lett. \textbf{71}, 4287 (1993)

\bibitem {swapping} J.-W. Pan, D. Boumweester, H. Weinfurter and A.
Zeilinger, Phys. Rev. Lett. \textbf{80}, 3891 (1998).

\bibitem {bennett} C. H. Bennett, \textit{et al.}, Phys. Rev.
Lett. \textbf{76}, 722 (1996).

\bibitem {repeater1}H.-J. Briegel \textit{et al.}, Phys. Rev. Lett.
\textbf{81}, 5932 (1998)

\bibitem {repeater2}W. D\"{u}r \textit{et al.}, Phys. Rev. A \textbf{59}, 169 (1999).

\bibitem {van}S.J. van Enk, J.I. Cirac, and P. Zoller, Science \textbf{279},
205 (1998).

\bibitem {DLCZ}L.-M. Duan, M. Lukin, J.I. Cirac, and P. Zoller, Nature
(London) \textbf{414}, 413 (2001).

\bibitem {nonclassical1}A. Kuzmich \textit{et al.}, Nature (London)
\textbf{423}, 731 (2003).

\bibitem {nonclassical2}C.H. van del Wal \textit{et al.}, Science
\textbf{301}, 196 (2003).

\bibitem {Eisman}M.D. Eisaman \textit{et al.}, Nature (London) \textbf{438},
837 (2005).

\bibitem {kuzmich}D.N. Matsukevich \textit{et al.}, Phys. Rev. Lett.
\textbf{\ 97}, 013601 (2006)

\bibitem {shuai}S. Chen \textit{et al.}, Phys. Rev. Lett. \textbf{97}, 173004 (2006).

\bibitem {entanglement}C.W. Chou \textit{et al.}, Nature (London)
\textbf{438}, 828 (2005).

\bibitem {kuzsci}D.N. Matsukevich and A. Kuzmich, Science \textbf{306}, 663 (2004).

\bibitem {segment}C.W. Chou \textit{et al.}, Science Express, 10.1126/science.1140300.

\bibitem {ye}K.W. Holman, D.D. Hudson, and J. Ye, Opt. Lett. \textbf{30}, 1225 (2005).

\bibitem {zhao}B. Zhao \textit{et. al.}, to appear in Phys. Rev. Lett (2007).

\bibitem {HOM}C.K. Hong, Z.Y. Ou, and L. Mandel, Phys. Rev. Lett. \textbf{59},
2044 (1987).

\bibitem {kimho}D. Felinto \textit{et al.}, Nature Physics \textbf{2}, 844 (2006).

\bibitem {kuzmho}T. Chaneli\`{e}re \textit{et al}., Phys. Rev. Lett.
\textbf{98}, 113602 (2007).

\bibitem {yuan}Z.-S. Yuan \textit{et al.}, Phys. Rev. Lett. \textbf{98} 180503 (2007).

\bibitem {thesis}A. Andr\'{e}, PhD thesis, Harvard University (2005).

\bibitem {jiang}L. Jiang \textit{et al.}, quant-ph/0609236.

\bibitem {Simon}C. Simon and W.T.M. Irvine, Phys. Rev. Lett. \textbf{91},
110405 (2003).

\bibitem {Feng}S. Bose \textit{et al}., Phys. Rev. Lett. \textbf{83}, 5158
(1999); X.-L. Feng \textit{et al}., \textit{ibid}. \textbf{90}, 217902 (2003);
D.E. Browne, M.B. Plenio, and S.F. Huelga, \textit{ibid}. \textbf{91}, 067901 (2003).

\bibitem {two-photon}D.E. Browne and T. Rudolph, Phys. Rev. Lett. \textbf{95},
010501 (2005).

\bibitem {chen}Y.-A. Chen \textit{et al.}, quant-ph/arXiv:0705.1256v1.

\bibitem {note}Note that in Ref. \cite{zhao}, both entnaglement swapping and
entanglement generation is performed remotely, so we need at least two atomic
ensembles at each node. Here entanglement swapping is performed locally
following the standard quantum repeater protocol \cite{repeater1}, and thus we
have to manipulate at least four atomic ensembles at each communication site.

\bibitem {optimal}A.V. Gorshkov \textit{et al.}, Phys. Rev. Lett. \textbf{98}, 123601.

\bibitem {lukin}M.D.\ Lukin, S.F. Yelin, and M. Fleischhauer, Phys. Rev. Lett.
\textbf{84}, 4232 (2000).

\bibitem {Fleis}M.\ Fleischhauer and M.D. Lukin, Phys. Rev. Lett. \textbf{84},
5094 (2000).

\bibitem {purify1}J.-W. Pan \textit{et al.}, Nature (London) \textbf{410},
1067 (2001)

\bibitem {purify2}J.-W. Pan \textit{et al.}, Nature (London) \textbf{423}, 417 (2003).

\bibitem {peng}C.-Z. Peng \textit{et al.}, Phys. Rev. Lett., \textbf{94},
150501 (2005).

\bibitem {kwait}D.F.V. James and P.G. Kwiat, Phys. Rev. Lett. \textbf{89},
183601 (2002)

\bibitem {imm}A. Imamo\={g}lu, Phys. Rev. Lett. \textbf{89}, 163602 (2002).

\bibitem {simon2}C. Simon, \textit{et al.}, Phys. Rev. Lett. \textbf{98},
190503 (2007).

\bibitem {simon3}C. Simon, private communicaiton.

\bibitem {hours}A. Dantan \textit{et al.}, Phys. Rev. Lett. \textbf{95},
123002 (2005).

\bibitem {free}T. Schmitt-Manderbach \textit{et al.}, Phys. Rev. Lett.
\textbf{98}, 010504 (2007).

\bibitem {peng2}C.Z. Peng \textit{et al.}, Phys. Rev. Lett. \textbf{98},
010505 (2007).

\bibitem {event}Q. Zhang \textit{et al.}, quant-ph/0610145.
\end{thebibliography}
\end{document}